\documentclass[12pt]{article}
\begin{document}

\title{ Comment on Phys. Rev. Lett. {\bf 110}, 168702 (2013): \\ "Causal Entropic Forces" }
\author{ Enrique Canessa\footnote{canessae@ictp.it} \\ 
{\small {\it The Abdus Salam International Centre for Theoretical Physics, Trieste, Italy } }  } 
\date{}
\maketitle

The recent Letter by Wissner-Gross and Freer \cite{Wis13} proposes a relationship between
intelligence and entropy maximization based on a causal generalization of entropic forces over configuration 
space paths, which may beautifully induce sophisticated behaviors associated with competitive adaptation on 
time scales faster than natural evolution.  These authors suggest a potentially general thermodynamic
model of adaptive behavior as a non-equilibrium process in open systems.
On the basis of the force-entropy correlations published by us a decade ago \cite{Can04}, we point out that 
their main relations have been previously reported within a simpler statistical thermodynamic model where 
non-interacting moving particles are assumed to form an elastic body.  

The claim in \cite{Wis13} that spontaneous emergence of adaptive behaviour (driven by the systems degrees of
freedom $j$ with internal Gaussian forces $f_{j}(t)$) maximizes the overall diversity of accessible future 
paths is only partially true.  There is an alternative approach to understand these complex networks 
as delineated from a probabilistic perspective within the canonical Gibbs distribution.

In our discretized formalism \cite{Can04}, the probability $p_{i}$ that the system is in the state $i=1, \dots N$, 
is given by two positive functions satisfying the normalization $\sum_{i=1}^{N} u_{i}w_{i} \equiv 1$.
This simple multiplicative form also gives interesting connections between an applied tension and thermodynamics
quantities of dynamical systems.  Such class of normalized product of positive functions for $p_{i}$
appears formally, {\it e.g.}, in the analysis of stochastic processes on graphs according to the Hammersley-Clifford Theorem.
As we have shown the product $u_{i}w_{i}$ leads to reveal intrinsic molecular-mechanical properties on classical 
and non-extensive dynamical systems in relation to a distinct tensile force acting on these systems at constant volume 
and number of particles with trajectories $x(t) = ({\bf q}(t),{\bf p}(t))$.  A new scenario for the entropic $q$-index in 
Tsallis statistics in terms of the energy of the system was also reported earlier --which has been applied to study, {\it e.g.}, brain dynamics.  

For completeness the causal entropic forces found in \cite{Wis13} and \cite{Can04} and derived from
rather alternative thermodynamic analytical models are listed below where the force $f_{i}$ represents 
variations in the energy states with respect to particle displacements.

\begin{table}[ht]
\centering
\begin{tabular}{|l|l|}
  \hline
   {\bf Continous Theory} \cite{Wis13}   & {\bf Discretized Theory}  \cite{Can04} \\
{\it A.D. Wissner-Gross and C.E. Free} & {\it E. Canessa}  \\[5ex]
  \hline
  $F_{j}( {\bf X_{o}},\tau ) =  T_{c} \; \frac{ \partial S_{c}({\bf X},\tau) }{ \partial q_{j}(0) }) |_{ {\bf X=X_{o}} }$        &   
  $F = k_{B}T \; \frac{\partial}{\partial x}
     \left( \frac{\sum_{i=1}^{N} p_{i}^{q}}{q-1} \right)_{T} = T \; \frac{\partial S_{q}}{\partial x}$
 \\[3ex] 
\hline
 $F_{j}( {\bf X_{o}},\tau ) = -k_{B}T \;\int_{_{{{\bf X}}(t)}} \frac{ \partial \Pr( {\bf x}(t) | {\bf x}(0) ) }{ \partial q_{j}(0) }
     \ln \Pr( {\bf x}(t) | {\bf x}(0) ) D {\bf x}(t)$   &  
$ F = - k_{B}T \; \sum_{i=1}^{N} (\frac{\partial p_{i}}{\partial x}) \ln p_{i}$
  \\[3ex] 
\hline
 $\frac{ \partial \Pr( {\bf x}(\epsilon) | {\bf x}(0) ) }{\partial q_{j}(0)} = \frac{2 f_{j}(0)}{k_{B}T} \Pr( {\bf x}(\epsilon) | {\bf x}(0) )$     &  
$\frac{\partial p_{i}}{\partial x} = - ( \frac{\partial \epsilon_{i}}{\partial x} ) \frac{p_{i}}{k_{B}T} \; \rightarrow
   \frac{f_{i}}{k_{B}T} \; p_{i}$
  \\[3ex] 
\hline
$F_{j}( {\bf X_{o}},\tau ) = - \frac{2 T_{c}}{T_{r}} \int_{_{{{\bf X}}(t)}} f_{j}(0)
  \Pr( {\bf x}(t) | {\bf x}(0) ) \ln  \Pr( {\bf x}(t) | {\bf x}(0) ) D {\bf x}(t)$    &
 $F \rightarrow \; - \sum_{i=1}^{N} f_{i} \; p_{i} \ln p_{i}$  
  \\[3ex] 
\hline
\end{tabular}
\end{table}


\begin{thebibliography}{99}

\bibitem{Wis13} A.D. Wissner-Gross and C.E. Free, Phys. Rev. Lett. {\bf 110} (2013) 168702
\bibitem{Can04} E. Canessa, Physica A {\bf 341} (2004) 165 -also at: arXiv:cond-mat/0403724

\end{thebibliography}
\end{document}